\documentclass[%
 reprint,
 amsmath,amssymb,
 aps,
]{revtex4-2}
 
\usepackage{xcolor}
\usepackage{graphicx}
\usepackage{dcolumn}
\usepackage{bm}

\begin{document}

\preprint{APS/123-QED}

\title{Thermodynamic potentials from a probabilistic view on the system-environment interaction energy}
\author{Mohammad Rahbar}
\affiliation{Technical University of Munich; TUM School of Natural Sciences, Department of Chemistry, Lichtenbergstr. 4, D-85748 Garching, Germany}
\author{Christopher J. Stein}%
\affiliation{Technical University of Munich; TUM School of Natural Sciences, Department of Chemistry, Catalysis Research Center, Atomistic Modeling Center, Munich Data Science Institute,Lichtenbergstr. 4, D-85748 Garching, Germany}
\email{christopher.stein@tum.de}

\date{\today}

\begin{abstract}
In open systems with strong coupling, the interaction energy between the system and the environment is significant, so thermodynamic quantities cannot be reliably obtained by traditional statistical mechanics methods. The
Hamiltonian of mean force $\mathcal{H}^{*}_{\beta}$ offers an in principle accurate theoretical basis by explicitly accounting for the interaction energy. However, calculating the Hamiltonian of mean force is challenging both
theoretically and computationally. We demonstrate that when the condition $\text{Var}_{\mathcal{E}_0} (e^{-\beta {V}_{\mathcal{SE}}}) = 0$ is met, the dependence of thermodynamic variables can be shifted from
$\{P_{\beta}(x_{\mathcal{S}}), \mathcal{H}^{*}_{\beta}(x_{\mathcal{S}})\}$ to $\{P_{\beta}(x_{\mathcal{S}}), P(V_{\mathcal{SE}})\}$. This change simplifies thermodynamic measurements. As a central result, we derive a general equality that holds for arbitrary coupling strengths and from which an inequality follows—aligned with Jensen’s inequality applied to the Gibbs–Bogoliubov–Feynman bound. This equality, analogous in importance to the Jarzynski equality, offers deeper insight into free energy differences in strongly coupled systems. Finally, by combining our result with said Jarzynski equality, we derive additional relations that further clarify thermodynamic behavior in strongly coupled open systems. 
\end{abstract}

\maketitle

The standard relations of statistical mechanics often break down for open systems that are strongly coupled to their environment. The main reason is that the interaction energy between system and environment can be comparable in magnitude to the system’s internal energy. In such cases, theories that go beyond conventional statistical mechanics are required to accurately describe thermodynamic behavior \cite{jarzynski2017stochastic,vilar2008failure,PhysRevLett.101.098901,PhysRevLett.101.098903,seifert2012stochastic,sekimoto1998langevin,zhang2012stochastic,van2015ensemble,esposito2010entropy,seifert2016first,talkner2016open,miller2017entropy,goldstein2019nonequilibrium,campa2009statistical,jarzynski2017stochastic,ford1985quantum,binder2018thermodynamics,talkner2020colloquium,thirring2003negative,hanggi1990reaction,cresser2021weak,ochoa2016energy,hilfer2022foundations,campisi2009fluctuation}. One widely accepted approach is the Hamiltonian of mean force (HMF), which incorporates system--environment interactions into an effective Hamiltonian for the system. The HMF formalism provides a consistent framework to define thermodynamic quantities such as internal energy and free energy by including an average interaction term alongside temperature \cite{feynman2000theory,kirkwood1935statistical,roux1999implicit,talkner2016open,talkner2020colloquium,jarzynski2017stochastic,miller2018energy,seifert2016first,miller2017entropy,talkner2016open,talkner2020colloquium,anto2023effective}. In practice, however, calculating the HMF is notoriously difficult because it requires evaluating a partition function that is tractable only in special cases. Additionally, recent works have highlighted a non-uniqueness in the definition of the HMF when inferred from the system’s probability density function (PDF) $P_{\beta}(x_\mathcal{S})$ \cite{talkner2016open,strasberg2020measurability,campisi2009fluctuation,talkner2020colloquium,xing2024thermodynamics, colla2025observing}. To illustrate the challenges arising in the HMF framework, we examine two cases in more detail. The first case concerns the challenge of determining the free energy. From the relation $-\beta^{-1} \ln P_{\beta}(x_\mathcal{S}) = \mathcal{H}_{\beta}^*(x_\mathcal{S}) - F^{*}_\mathcal{S}(\beta)$ it is easy to see that for a given PDF, which uniquely represents the statistical behavior of the system, no unique free energy can be obtained \cite{talkner2016open}. This is because the equation can be satisfied for infinitely many choices of the $\mathcal{H}_{\beta}^*(x_\mathcal{S})$ and $F^{*}_\mathcal{S}(\beta)$ pair. In simpler terms, knowing the statistical behavior of the system and using it as an physical observable in the above equation leads to an equation with two unknowns. The second case is the calculation of the system's internal energy, $U_S$. In the HMF framework, we have  $U_S = \langle \partial (\beta \mathcal{H}_{\beta}^*(x_\mathcal{S}))/\partial \beta \rangle_{\mathcal{S}}$, where
$\langle \bullet \rangle_{\mathcal{S}} = \int \mathrm{d} x_\mathcal{S} \, \bullet \, P_{\beta}(x_\mathcal{S})$. The PDF is $P_{\beta}(x_\mathcal{S}) = \mathcal{Z}_{*}^{-1}(\beta) \exp(-\beta \mathcal{H}_{\beta}^*(x_\mathcal{S}))$, and the partition function is $\mathcal{Z}_{*}(\beta) = \int \mathrm{d} x_\mathcal{S} \exp(-\beta \mathcal{H}_{\beta}^*(x_\mathcal{S}))$. As seen from the above relation, even if one can precisely calculate the PDF and HMF by some experimental or theoretical means and substitute them into that relation, the internal energy still cannot be computed. The reason is that the temperature dependence of the HMF leads to the appearance of a term involving $\beta \langle {\partial \mathcal{H}_{\beta}^*(x_\mathcal{S})}/{\partial \beta} \rangle_{\mathcal{S}}$ in the calculation of the internal energy \cite{talkner2020colloquium}. Therefore, determining the internal energy requires precise understanding of how the HMF depends on temperature, which is very complicated and often impossible in practice \cite{talkner2016open,talkner2020colloquium}. In light of these issues, it is desirable to reformulate the thermodynamics of strongly coupled systems in terms of more accessible quantities. In this work, we demonstrate that under a certain condition, one can avoid direct use of the HMF. We show that one can shift the dependence of the thermodynamic variables from  $\{P_{\beta}(x_{\mathcal{S}}), \mathcal{H}^{*}_{\beta}(x_{\mathcal{S}})\}$ to $
\{P_{\beta}(x_{\mathcal{S}}), P(V_{\mathcal{SE}})\}$
under the condition $\mathrm{Var}_{\mathcal{E}_0}(e^{-\beta V_{\mathcal{SE}}}) = 0$.\\
To arrive there, we begin with the equation   
\begin{equation}
\exp\Bigr( -\beta \mathcal{H}_{\beta}^{*}(x_\mathcal{S}) + \beta \mathcal{H}_{\mathcal{S}}(x_\mathcal{S}) \Bigr) =  \Bigr\langle \exp\left( -\beta {V}_{\mathcal{S}\mathcal{E}} \right) \Bigr\rangle_{\mathcal{E}_0}
\label{asli},
\end{equation}
where $\langle \cdot \rangle_{\mathcal{E}_0}$ denotes the average computed using the probability distribution, $P_{\mathcal{E}_0}(x_{\mathcal{E}})$, defined over the environment's degrees of freedom \cite{talkner2016open,talkner2020colloquium}. It is important to mention that this PDF assigned to the unperturbed environment is free from any influence of the system. Therefore, we can write  $P_{\mathcal{E}_0}(x_{\mathcal{E}}) = Z_{\mathcal{E}}^{-1} \exp(-\beta \mathcal{H}_{\mathcal{E}}(x_{\mathcal{E}}))$, 
where the environmental partition function is  $Z_{\mathcal{E}} = \int \mathrm{d}x_{\mathcal{E}}\, \exp(-\beta \mathcal{H}_{\mathcal{E}}(x_{\mathcal{E}}))$. 
We next average over the system PDF --- abbreviated as $\langle \bullet \rangle_{\mathcal{S}}$, where $\bullet$ is a placeholder for any observable --- on both sides of Eq.~(\ref{asli}) to obtain  
\begin{equation}
\Bigr\langle\exp\Bigr( -\beta \mathcal{H}_{\beta}^{*}(x_\mathcal{S}) + \beta \mathcal{H}_{\mathcal{S}}(x_\mathcal{S}) \Bigr) \Bigr\rangle_{\mathcal{S}} = \Bigr\langle \Bigr\langle \exp\left( -\beta {V}_{\mathcal{S}\mathcal{E}} \right) \Bigr\rangle_{\mathcal{E}_0} \Bigr\rangle_{\mathcal{S}}.
\label{eq:HMF}
\end{equation}
By substituting $\langle \langle \exp\left( -\beta {V}_{\mathcal{S}\mathcal{E}} \right) \rangle_{\mathcal{E}_0} \rangle_{\mathcal{S}}$ with $\langle\exp\left( -\beta {V}_{\mathcal{S}\mathcal{E}} \right)\rangle_{\mathcal{S+E}}$ we will ultimately eliminate the free energy’s dependence on the HMF. In this substitution,  $\left\langle \bullet \right\rangle _{\mathcal{S}+\mathcal{E}}$ is defined as  
\begin{align}
\left\langle \bullet \right\rangle _{\mathcal{S}+\mathcal{E}} &= \int \mathrm{d} x_\mathcal{S} \int \mathrm{d} x_\mathcal{E} \, \bullet \, P(x_\mathcal{S}, x_\mathcal{E}) \nonumber \\
&= \int \mathrm{d} x_\mathcal{S} \int \mathrm{d} x_\mathcal{E} \, \bullet \, \frac{\exp\left( -\beta \mathcal{H}_{\mathcal{S}+\mathcal{E}}(x_\mathcal{S}, x_\mathcal{E}) \right)}{Z_{\mathcal{S}+\mathcal{E}}}.
\label{eq:total_average}
\end{align}  
Effectively, this substitution achieves the desired variable transformation that we mentioned earlier.
An important issue here is to determine under what condition that substitution is permitted. This is identical to finding a condition that leads to the equality $ \langle\exp (-\beta {V}_{\mathcal{SE}})\rangle_{\mathcal{E}_0}= \langle\exp (-\beta {V}_{\mathcal{SE}})\rangle_{\mathcal{E}|\mathcal{S}}$. The conditional expectation of an observable, $\langle \bullet \rangle_{\mathcal{E}|\mathcal{S}}$, under a fixed degree of freedom of the system can be written as
\begin{equation}
\langle \bullet \rangle_{\mathcal{E}|\mathcal{S}} = \int \mathrm{d}x_\mathcal{E} \, \bullet \, P(x_\mathcal{E} \mid x_\mathcal{S}),
\label{eq:conditional_average}
\end{equation}  
where the corresponding conditional probability distribution is \cite{talkner2016open}
\begin{equation}
P(x_\mathcal{E} \mid x_\mathcal{S}) =\frac{1}{Z_{\mathcal{E}}} \exp\Bigl( -\beta \mathcal{H}_{\mathcal{S}+\mathcal{E}}(x_\mathcal{S}, x_\mathcal{E}) + \beta \mathcal{H}^{*}(x_\mathcal{S}) \Bigr).
\label{eq:conditional_probability}
\end{equation}
Using the definitions above and $\mathcal{H}_{\mathcal{S}+\mathcal{E}}(x_\mathcal{S}, x_\mathcal{E}) = \mathcal{H}_{\mathcal{S}}(x_\mathcal{S}) + \mathcal{H}_{\mathcal{E}}(x_\mathcal{E})+ V(x_\mathcal{S},x_\mathcal{E})$, one can write
\begin{align}
&\nonumber\Bigr\langle\exp(-\beta V(x_\mathcal{S},x_\mathcal{E}))\Bigr\rangle_{\mathcal{E}|\mathcal{S}}
\\&\nonumber=\exp\Bigr(\beta\mathcal{H}_\beta^*(x_s)\Bigr)\exp\Bigr(-\beta\mathcal{H}_{\mathcal{S}}(x_s)\Bigr) \\&\times\int\mathrm{d}x_\mathcal{E} \frac{\exp(-2\beta V(x_\mathcal{S},x_\mathcal{E}))\exp(-\beta\mathcal{H}_\mathcal{E}(x_\mathcal{E}))}{\mathcal{Z_E}}.
\end{align}  
Using Eq.~(\ref{asli}), the prefactor outside the integral on the right-hand side can be replaced by $\langle e^{-\beta V_{\mathcal{S}\mathcal{E}}} \rangle^{-1}_{\mathcal{E}_0}$, while the integral itself evaluates to $\langle e^{-2\beta V_{\mathcal{S}\mathcal{E}}} \rangle_{\mathcal{E}_0}$. We hence obtain
\begin{align}
\Bigr\langle\exp (-\beta {V}_{\mathcal{SE}})\Bigr\rangle_{\mathcal{E}|\mathcal{S}} = \frac{\Bigr\langle\exp(-2\beta {V}_{\mathcal{SE}})\Bigr\rangle_{\mathcal{E}_0}}{\Bigr\langle\exp(-\beta {V}_{\mathcal{SE}})\Bigr\rangle_{\mathcal{E}_0}}.
\end{align}
From this equation, we see that the condition $\langle e^{-\beta V_{\mathcal{S}\mathcal{E}}} \rangle_{\mathcal{E}_0} = \langle e^{-\beta V_{\mathcal{S}\mathcal{E}}} \rangle_{\mathcal{E}|\mathcal{S}}$ is satisfied only when
\begin{equation}
 \Bigr\langle\exp(-2\beta {V}_{\mathcal{SE}})\Bigr\rangle_{\mathcal{E}_0} -\Bigr\langle\exp(-\beta {V}_{\mathcal{SE}})\Bigr\rangle_{\mathcal{E}_0}^2 = 0. 
\label{eq: variance}
\end{equation}
This is equivalent to writing $\text{Var}_{\mathcal{E}_0} (e^{-\beta {V}_{\mathcal{SE}}}) = 0$.\\
We now discuss several physically meaningful scenarios for the system and environment coupling that satisfy Eq.~(\ref{eq: variance}). 
The most obvious case is when $V(x_\mathcal{S},x_\mathcal{E}) = 0$ or $V(x_\mathcal{S},x_\mathcal{E}) \approx 0$, which corresponds to non-existing or very weak system-environment coupling. In addition to that almost trivial case, we can also describe two specific strongly coupled scenarios that fulfill this condition. One case is when we assume $V(x_\mathcal{S},x_\mathcal{E}) = g^b(x_\mathcal{S};\mathcal{E}^b)$, meaning that the interaction between the system and the environment depends only on the state of the system and on macroscopic variables of the surrounding environment, $\mathcal{E}^b$, that are unaffected by the system’s state.
In other words, the environment acts as a fixed external field for the system. 
Another case is $V(x_\mathcal{S},x_\mathcal{E}) = g^*(x_\mathcal{S},x_\mathcal{E}^*(x_\mathcal{S}))$, in which the interaction between the system and the environment depends on both the system’s state and the environment’s state, with the latter instantaneously adapting to the system’s state. In that scenario, a time-scale difference is required between the change of the system's coordinates and the sudden response of the environment (water molecules adapting to a protein's large amplitude motions would be an adequate example for this scenario) \cite{mondal2017origin, fogarty2014water}.\\
One important point to address is whether the substitution is consistent with the definition of the partition function in the HMF framework. In other words, whether by accepting the following equation
\begin{align}
\mathcal{H}_\beta^{*}(x_\mathcal{S}) = \mathcal{H}_{\mathcal{S}}(x_\mathcal{S}) - \beta^{-1} \ln \Bigr\langle \exp(-\beta {V}_{\mathcal{S}\mathcal{E}}) \Bigr\rangle_{\mathcal{E|S}}.
\label{Jaygozin}
\end{align}  
under the condition $\text{Var}_{\mathcal{E}_0} (e^{-\beta {V}_{\mathcal{SE}}}) = 0$ we can obtain $\mathcal{Z}_* = \mathcal{Z}_{\mathcal{S+E}} /\mathcal{Z}_{\mathcal{E}}$ \cite{burke2024structure}. This relation is the most fundamental assumption in HMF theory, as all thermodynamic relations in the HMF framework depend on it. To check the consistency, we rewrite Eq.~(\ref{Jaygozin}) to yield 
\begin{align}
&\nonumber\exp(-\beta\mathcal{H}_{\beta}^{*}(x_{\mathcal{S}})) \\&=  \int \mathrm{d} x_{\mathcal{E}} \exp(-\beta\mathcal{H}_{\mathcal{S}}(x_\mathcal{S}))\exp (-\beta{V}_{\mathcal{SE}}) P(x_\mathcal{E}|x_\mathcal{S}).
\end{align}
By multiplying both sides of the above equation by $P_\beta(x_\mathcal{S})/\mathcal{Z}_{\mathcal{E}}$, substituting $P(x_\mathcal{E}|x_\mathcal{S})\, P_\beta(x_\mathcal{S})$ on the right-hand side with $P(x_\mathcal{S},x_\mathcal{E}) $, and by using Eq.~(\ref{eq:total_average})  and Eq.~(\ref{Jaygozin}),  we obtain after simplification
\begin{align}
\frac{\mathcal{Z}_{\mathcal{S+E}}}{\mathcal{Z}^*\mathcal{Z}_{\mathcal{E}}} = \frac{\left\langle\exp(-2\beta{V}_{\mathcal{SE}})\right\rangle_{\mathcal{E}_0}}{\left\langle\exp(-\beta{V}_{\mathcal{SE}})\right\rangle^2_{\mathcal{E|S}}}.
\end{align}
Now, by applying the condition $\text{Var}_{\mathcal{E}_0} (e^{-\beta {V}_{\mathcal{SE}}}) = 0$ on the right-hand side of the above equation, it follows that the right-hand side equals 1. Consequently, we obtain the fundamental definition of the partition function in the HMF framework.\\
By accepting the equality of $\langle \langle \exp\left( -\beta {V}_{\mathcal{S}\mathcal{E}} \right) \rangle_{\mathcal{E}_0} \rangle_{\mathcal{S}}$ with $\langle\exp\left( -\beta {V}_{\mathcal{S}\mathcal{E}} \right)\rangle_{\mathcal{S+E}}$ under the above condition we can substitute the right-hand side of equation Eq.~(\ref{eq:HMF}) with $\langle\exp\left( -\beta {V}_{\mathcal{S}\mathcal{E}} \right)\rangle_{\mathcal{S+E}}$, and write
\begin{equation}
\Bigr\langle\exp\Bigr( -\beta \mathcal{H}_{\beta}^{*}(x_\mathcal{S}) + \beta \mathcal{H}_{\mathcal{S}}(x_\mathcal{S}) \Bigr) \Bigr\rangle_{\mathcal{S}} = \Bigr\langle \exp(-\beta {V}_{\mathcal{S}\mathcal{E}}) \Bigr\rangle_{\mathcal{S}+\mathcal{E}}.
\label{HMF_var}
\end{equation}
After expressing $\mathcal{H}_\beta^{*}(x_\mathcal{S})$ in terms of $P_\beta(x_\mathcal{S})$ and $F^*_{\mathcal{S}}$ (\textit{vide supra}) and simplifying the resulting expression, one obtains
\begin{equation}
\exp( -\beta F^*_{\mathcal{S}})  = \frac{\Bigr\langle \exp(-\beta {V}_{\mathcal{S}\mathcal{E}}) \Bigr\rangle_{\mathcal{S}+\mathcal{E}}}{\Bigr\langle P_\beta(x_\mathcal{S}) \exp\Bigr( \beta \mathcal{H}_{\mathcal{S}}(x_\mathcal{S}) \Bigr) \Bigr\rangle_{\mathcal{S}}}.
\label{eq:free_energy_relation}
\end{equation}  
As shown in the above equation, which holds under the condition $\text{Var}_{\mathcal{E}_0}(e^{-\beta V_{\mathcal{SE}}}) = 0$, the free energy no longer depends on the HMF as in $\{P_{\beta}(x_{\mathcal{S}}), \mathcal{H}^*_{\beta}(x_{\mathcal{S}})\}$ but instead becomes fully determined by the joint input $\{P_{\beta}(x_{\mathcal{S}}), P_{\beta}(V_{\mathcal{S}\mathcal{E}})\}$.\\
Under the same condition, the internal energy can be expressed with these new variables. 
To show this, we start with the following definition of the internal energy \cite{talkner2016open,talkner2020colloquium}:
\begin{equation}
U_{\mathcal{S}} = \left\langle \mathcal{H}_{\mathcal{S}} \right\rangle_{\mathcal{S}} + \left\langle {V}_{\mathcal{S}\mathcal{E}} \right\rangle_{\mathcal{S}+\mathcal{E}} + \left\langle \mathcal{H}_{\mathcal{E}} \right\rangle_{\mathcal{S}+\mathcal{E}} - \left\langle \mathcal{H}_{\mathcal{E}} \right\rangle_{\mathcal{E}_0}.
\end{equation}
Next, we prove that under the variance condition one can conclude that
\begin{equation}
\left\langle \mathcal{H}_{\mathcal{E}} \right\rangle_{\mathcal{S}+\mathcal{E}} - \left\langle \mathcal{H}_{\mathcal{E}} \right\rangle_{\mathcal{E}_0}=0.
\label{eq_15}
\end{equation}
To show this, we begin with the term $\left\langle \mathcal{H}_{\mathcal{E}} \right\rangle_{\mathcal{S}+\mathcal{E}}$ Using the definition established in Eq.~(\ref{eq:total_average}) and also substituting the partition function relation $\mathcal{Z}^* = \mathcal{Z}_{\mathcal{S+E}} /\mathcal{Z}_{\mathcal{E}}$, we have
\begin{align}
&\nonumber\langle\mathcal{H}_\mathcal{E}\rangle_\mathcal{S+E}=\int \mathrm{d}x_\mathcal{S}\int\mathrm{d}x_\mathcal{E} \mathcal{H}_\mathcal{E} P (x_\mathcal{S},x_\mathcal{E})\\
&= \int \mathrm{d}x_\mathcal{E} \mathcal{H}_\mathcal{E} P_{\mathcal{E}_0}(x_{\mathcal{E}}) \underbrace{\int \mathrm{d}x_\mathcal{S} \frac{\exp(-\beta \mathcal{H}_\mathcal{S} )\exp(-\beta {V}_{\mathcal{S}\mathcal{E}})}{\mathcal{Z}^*}}_{M(x_\mathcal{E})}.
\end{align}
In the above expression, the function $M(x_\mathcal{E})$ can be rewritten using Eq.~(\ref{asli}) or Eq.~(\ref{Jaygozin}) as follows
\begin{equation}
M(x_\mathcal{E}) = \int \mathrm{d}x_\mathcal{S}\frac{\exp(-\beta\mathcal{H}^{*}_\beta(x_{\mathcal{S}}))}{\mathcal{Z}^*}\frac{\exp(-\beta {V}_{\mathcal{S}\mathcal{E}})}{\Bigr\langle \exp\left( -\beta {V}_{\mathcal{S}\mathcal{E}} \right) \Bigr\rangle_{\mathcal{E}_0}}.
\end{equation}
It is straightforward to verify that, in each of the three scenarios derived from the variance condition, ${\exp(-\beta {V}_{\mathcal{S}\mathcal{E}})}/{\langle \exp\left( -\beta {V}_{\mathcal{S}\mathcal{E}} \right) \rangle_{\mathcal{E}_0}}$ =1. Due to the probability property, the remainder of the integral becomes equal to one and Eq. (\ref{eq_15}) holds. 
In this way, one arrives at $U_{\mathcal{S}} = \left\langle \mathcal{H}_{\mathcal{S}} \right\rangle_{\mathcal{S}} + \left\langle V_{\mathcal{S}\mathcal{E}} \right\rangle$ for calculating internal energy.\\
In the following, we will show that from Eq.~(\ref{eq:HMF}) and a specific interpretation of the Free Energy Perturbation (FEP) method \cite{kirkwood1935statistical, zwanzig1954high}, one can arrive at an inequality that is always valid regardless of the coupling strength. Theoretically, the FEP method is accurate for any perturbation \cite{chipot2007free}. 
Using FEP, the free energy difference between a reference state and a target state can be expressed as follows 
\begin{equation}
e^{-\beta (F_{\mathcal{S+E}}-F_\mathcal{S}-F_\mathcal{E})} = \Bigr\langle e ^{-\beta V_{\mathcal{SE}}}\Bigr\rangle_{\mathcal{S}_0+\mathcal{E}_0} =\Bigr\langle e ^{\beta V_{\mathcal{SE}}}\Bigr\rangle_{\mathcal{S}+\mathcal{E}}^{-1},
\label{FEP_0}
\end{equation}
where $\langle \cdot\rangle_{\mathcal{S}_0+\mathcal{E}_0}$ is defined under the probability distribution $P_{\mathcal{S}_0} (x_\mathcal{S}) P_{\mathcal{E}_0}(x_{\mathcal{E}})$,  
with $P_{\mathcal{S}_0} (x_\mathcal{S}) = \exp(-\beta\mathcal{H}(x_\mathcal{S})) / \mathcal{Z_S}$. To use the above definition in the present discussion, we assume that in the target state  $F_{\mathcal{S+E}}$, the system and environment are strongly coupled and have reached equilibrium with their surrounding thermal bath. 
On the other hand, in the reference state $F_\mathcal{S}+F_\mathcal{E}$, we assume that the system and environment are completely isolated from each other and in equilibrium with a thermal bath. 
From the key definition of the partition function in the HMF framework, one can see $F_{\mathcal{S+E}} - F_{\mathcal{E}} = F^{*}_\mathcal{S}$ \cite{talkner2016open,talkner2020colloquium,PRE}. Now, using the interpretation we discussed for FEP and replacing $\mathcal{H}_\beta^{*}(x_\mathcal{S})$ with an expression in terms of $P_\beta(x_\mathcal{S})$ and $F^*_{\mathcal{S}}$, we can write
\begin{equation}
\exp( -\beta F^*_{\mathcal{S}})  = \frac{\Bigr\langle \exp(-\beta {V}_{\mathcal{S}\mathcal{E}})\Bigr\rangle_{\mathcal{S}_0+\mathcal{E}_0}}{ P_{\mathcal{S}_0}(x_\mathcal{S}) \exp\Bigr( \beta \mathcal{H}_{\mathcal{S}}(x_\mathcal{S}) \Bigr) },
\label{fep}
\end{equation}
Similarly, with Eq.~(\ref{eq:HMF}) we have
\begin{equation}
\exp( -\beta F^*_{\mathcal{S}})  = \frac{\Bigr\langle \Bigr\langle \exp\left( -\beta {V}_{\mathcal{S}\mathcal{E}} \right) \Bigr\rangle_{\mathcal{E}_0} \Bigr\rangle_{\mathcal{S}}}{\Bigr\langle P_\beta(x_\mathcal{S}) \exp\Bigr( \beta \mathcal{H}_{\mathcal{S}}(x_\mathcal{S}) \Bigr) \Bigr\rangle_{\mathcal{S}}}.
\label{free energyHMF}
\end{equation} 
If we set the right-hand sides of these two equations equal, we get
\begin{align}
 \frac{\Bigr\langle \exp(-\beta {V}_{\mathcal{S}\mathcal{E}})\Bigr\rangle_{\mathcal{S}_0+\mathcal{E}_0}}
{\Bigr\langle \Bigr\langle \exp\left( -\beta {V}_{\mathcal{S}\mathcal{E}} \right) \Bigr\rangle_{\mathcal{E}_0} \Bigr\rangle_{\mathcal{S}}} = \frac{ P_{\mathcal{S}_0}(x_\mathcal{S}) \exp\Bigr( \beta \mathcal{H}_{\mathcal{S}}(x_\mathcal{S}) \Bigr) }{\Bigr\langle P_\beta(x_\mathcal{S}) \exp\Bigr( \beta \mathcal{H}_{\mathcal{S}}(x_\mathcal{S}) \Bigr) \Bigr\rangle_{\mathcal{S}}}.
\label{sp-eq}
\end{align}  
Using Jensen's inequality \cite{jensen1906fonctions} from probability theory and the definition of $\langle \bullet \rangle_{\mathcal{S}}$, one can show that the right-hand side is less than or equal to one. This leads us to
\begin{align}
\Bigr\langle \exp(-\beta {V}_{\mathcal{S}\mathcal{E}})\Bigr\rangle_{\mathcal{S}_0+\mathcal{E}_0} \le  \Bigr\langle \Bigr\langle \exp\left( -\beta {V}_{\mathcal{S}\mathcal{E}} \right) \Bigr\rangle_{\mathcal{E}_0} \Bigr\rangle_{\mathcal{S}}.
\end{align}
Next, by using Eq.~(\ref{FEP_0}) and Eq.~(\ref{eq:HMF}), which come directly from the FEP theory and the HMF method, we reach the following inequality
\begin{align}
\exp( -\beta \Delta F_{\mathcal{S}}) \le \Bigr\langle\exp(-\beta\Delta\mathcal{H}_\beta(x_\mathcal{S}))\Bigr\rangle_{\mathcal{S}},
\label{Main}
\end{align}
where $\Delta F_{\mathcal{S}} = F^*_{\mathcal{S}} - F_{\mathcal{S}}$ and $\Delta\mathcal{H}_\beta(x_\mathcal{S}) = \mathcal{H}_\beta^*(x_\mathcal{S}) - \mathcal{H}(x_\mathcal{S})$. 
It is important to note that an equivalent form of Eq.~(\ref{Main}) can be obtained by applying Jensen’s inequality to the Gibbs–Bogoliubov–Feynman (GBF) inequality \cite{zhang1996application, faussurier2003gibbs, feynman2018statistical,shaw1986approximate,reible2023finite,delle2017partitioning}. 
Inspired by the inequality Eq.~(\ref{Main}), we seek a stronger statement—an exact equality—from which the inequality can be recovered as a corollary. 
We write the term $\exp(-\beta\Delta\mathcal{H}_\beta(x_\mathcal{S}))$ on the right-hand side of Eq.~(\ref{Main}) as \begin{align}
\exp(-\beta\Delta\mathcal{H}_\beta(x_\mathcal{S})) =\frac{P_\beta(x_\mathcal{S})\mathcal{Z^*}}{P_{\mathcal{S}_0}(x_\mathcal{S})\mathcal{Z}_\mathcal{S}}.
\end{align}
We take the average $\langle \bullet \rangle_{\mathcal{S}}$ on both sides, substitute the partition functions corresponding to the free energies, and simplify the expression to yield
\begin{align}
\frac{\Bigr\langle\exp(-\beta\Delta\mathcal{H}_\beta(x_\mathcal{S}))\Bigr\rangle_{\mathcal{S}}}{\exp( -\beta \Delta F_{\mathcal{S}})} = \int\mathrm{d}x_\mathcal{S} \frac{P^2_\beta(x_\mathcal{S})}{P_{\mathcal{S}_0}(x_\mathcal{S})}.
\label{Strong Equality}
\end{align}
One can show that the right-hand side of Eq.~(\ref{Strong Equality}) is always greater than or equal to $1$, which leads to Eq.~(\ref{Main}).
To see this explicitly, we recall Hölder's inequality \cite{holder1889ueber}, which states that for functions $f(x)$ and $g(x)$, and for exponents $p$ and $q$ that satisfy the condition, $\frac{1}{p} + \frac{1}{q} = 1$, the following relation holds
\begin{equation}
\int_S |f(x) g(x)| \mathrm{d}x \leq \left(\int_S |f(x)|^p \mathrm{d}x\right)^{\frac{1}{p}} \left(\int_S |g(x)|^q \mathrm{d}x\right)^{\frac{1}{q}}.
\end{equation}
If we choose $f(x) = \sqrt{P_{\mathcal{S}_0}}$ and $g(x) = {P_\beta}/{\sqrt{P_{\mathcal{S}_0}}}$, and also set $p=q=2$ we have
\begin{align}
&\nonumber\int\mathrm{d}x_\mathcal{S} \sqrt{P_{\mathcal{S}_0}} \frac{P_\beta}{\sqrt{P_{\mathcal{S}_0}}} \\&\leq \Bigr(\int\mathrm{d}x_\mathcal{S} \Bigr(\frac{P_\beta}{\sqrt{P_{\mathcal{S}_0}}}\Bigr)^2\Bigr)^\frac{1}{2} \Bigr(\int\mathrm{d}x_\mathcal{S} \Bigr(\sqrt{P_{\mathcal{S}_0}}\Bigr)^2\Bigr)^\frac{1}{2}.
\end{align}
Applying the normalization conditions
\begin{equation}
\int \mathrm{d} x_\mathcal{S} P_{\mathcal{S}_0}(x_\mathcal{S}) =1, \quad \int \mathrm{d} x_\mathcal{S} P_\beta(x_\mathcal{S})=1, 
\end{equation}
we obtain
\begin{align}
1 \leq \Bigr(\int\mathrm{d}x_\mathcal{S}\frac{P^2_\beta(x_\mathcal{S})}{P_{\mathcal{S}_0}(x_\mathcal{S})}\Bigr)^\frac{1}{2}.
\label{Hold}
\end{align}
A key finding of this work is obtained when introducing the definition of the chi-squared divergence ${\chi^2(P_\beta \parallel P_{\mathcal{S}_0})}$, which quantifies the difference between $P_\beta(x_\mathcal{S})$ and $P_{\mathcal{S}_0}(x_\mathcal{S})$ \cite{nishiyama2020relations}
\begin{align}
\chi^2\left(P_\beta \| P_{\mathcal{S}_0}\right)=-1+\int \mathrm{d} x_\mathcal{S} \Bigr(\frac{P_\beta(x _\mathcal{S})}{P_{\mathcal{S}_0}(x_\mathcal{S})}\Bigr)^2 P_{\mathcal{S}_0}(x_\mathcal{S}),
\label{chi}
\end{align}  
into Eq.~(\ref{Strong Equality}).
This leads to 
\begin{align}
e^{-\beta \Delta F_S}=\frac{\left\langle e^{-\beta \Delta \mathcal{H}_\beta(x)}\right\rangle_\mathcal{S}}{1+\chi^2\left(P_\beta \parallel P_{\mathcal{S}_0}\right)} . \label{General}
\end{align}   
Next, we will show how using Eq.~(\ref{General}) and combining it with existing theories like FEP leads to further insights into strongly coupled open systems. 
With the same definitions as in FEP (cf. Eq.~(\ref{FEP_0})), we can write \cite{zwanzig1954high,chipot2007free}
\begin{equation}
 e^{\beta \Delta F_S} = \left\langle e^{\beta \Delta \mathcal{H}_\beta(x)}\right\rangle_\mathcal{S} =\Bigr\langle e ^{-\beta V_{\mathcal{SE}}}\Bigr\rangle_{\mathcal{S}_0+\mathcal{E}_0}^{-1} =\Bigr\langle e ^{\beta V_{\mathcal{SE}}}\Bigr\rangle_{\mathcal{S}+\mathcal{E}}
 \label{FEP-like}
\end{equation}
Combining Eq.~(\ref{General}) and Eq.~(\ref{FEP-like}) gives
\begin{equation}
\Bigr\langle e^{\beta \Delta \mathcal{H}_\beta(x)}\Bigr\rangle_{\mathcal{S}} \Bigr\langle e^{-\beta \Delta \mathcal{H}_\beta(x)}\Bigr\rangle_{\mathcal{S}} = 1+\chi^2\left(P_\beta \parallel P_{\mathcal{S}_0}\right)  .
 \label{combine}
\end{equation}
It is important to emphasize that Eq.~(\ref{General}) and Eq.~(\ref{combine}) hold for any coupling strength. 
If the coupling is very weak, $P_\beta(x _\mathcal{S})\approx P_{\mathcal{S}_0}(x_\mathcal{S})$ \cite{talkner2016open,talkner2020colloquium,huang2022nonperturbative,campisi2009fluctuation}, so $\chi^2(P_\beta\|P_{\mathcal{S}_0})\approx 0$, and Eq.~(\ref{General}) reduces to $\langle e^{-\beta \Delta \mathcal{H}_\beta(x)}\rangle_\mathcal{S} = e^{-\beta \Delta F_S}$, which is trivially satisfied (since $\Delta \mathcal{H}_\beta(x)\approx 0$ and $\Delta F_S\approx 0$ in that limit).
In the cases discussed above where $\text{Var}_{\mathcal{E}_0}(e^{-\beta V_{\mathcal{SE}}}) = 0$ holds, by placing the result of Eq.~(\ref{HMF_var}) and Eq.~(\ref{FEP-like}) into Eq.~(\ref{combine}), we obtain
\begin{align}
& \Bigr\langle e^{\beta {V}_{\mathcal{S}\mathcal{E}}}\Bigr\rangle_{\mathcal{S+E}}\Bigr\langle e^{-\beta {V}_{\mathcal{S}\mathcal{E}} }\Bigr\rangle_{\mathcal{S+E}}  =1 +\chi^2\left(P_\beta \parallel P_{\mathcal{S}_0}\right)  .
\label{fep1}
\end{align}
On the other hand, from FEP we can prove \cite{chipot2007free}
\begin{align}
\Bigr\langle e^{-\beta {V}_{\mathcal{S}\mathcal{E}} }\Bigr\rangle_{\mathcal{S+E}} = \Bigr\langle e^{-\beta {V}_{\mathcal{S}\mathcal{E}} }\Bigr\rangle_{\mathcal{S}_0+\mathcal{E}_0}^{-1}\Bigr\langle e^{-2\beta {V}_{\mathcal{S}\mathcal{E}} }\Bigr\rangle_{\mathcal{S}_0+\mathcal{E}_0}.
\end{align}
With this and by inserting Eq.~(\ref{FEP-like}) into Eq.~(\ref{fep1}), we can equivalently write
\begin{align}
\Bigr\langle e^{-\beta {V}_{\mathcal{S}\mathcal{E}} }\Bigr\rangle_{\mathcal{S}_0+\mathcal{E}_0}^{-2}\Bigr\langle e^{-2\beta {V}_{\mathcal{S}\mathcal{E}} }\Bigr\rangle_{\mathcal{S}_0+\mathcal{E}_0} =1 +\chi^2\left(P_\beta \parallel P_{\mathcal{S}_0}\right).
\label{fep2}
\end{align}
To understand the relevance of Eq.~(\ref{fep1}) and Eq.~(\ref{fep2}) we assume that, in addition to the condition $ \text{Var}_{\mathcal{E}_0}(e^{-\beta V_{\mathcal{SE}}}) = 0 $, the interaction energy $ V_{\mathcal{SE}} $ is Gaussian distributed. The validity of this assumption can be argued from the central limit theorem but even if in practice the distribution $ P(V_{\mathcal{SE}}) $ is not exactly Gaussian, it usually has a Gaussian-like shape \cite{chipot2007free,naleem2017gaussian,park2004calculating,gore2003bias,gupta2020work}. From Eq.~(\ref{fep1}), it follows that
\begin{align}
\exp\left(\frac{\beta^2 \sigma^2}{2}\right) = \Bigl(  \chi^2\left(P_\beta \parallel P_{\mathcal{S}_0}\right) +1 \Bigr)^{0.5}, \label{sigma}
\end{align}
where $ \sigma^2 = \langle V_{\mathcal{S}\mathcal{E}}^2 \rangle_{\mathcal{S}+\mathcal{E}} - \langle V_{\mathcal{S}\mathcal{E}} \rangle_{\mathcal{S}+\mathcal{E}}^2 $ represents the variance of the interaction energy in the target setup, where the
system and environment are fully coupled. 
Applying the same reasoning for Eq.~(\ref{fep2}), we obtain  
\begin{align}
\exp\left(\frac{\beta^2 \sigma_0^2}{2}\right) =\Bigr(  \chi^2\left(P_\beta \parallel P_{\mathcal{S}_0}\right) +1 \Bigr)^{0.5}\label{sigma0},
\end{align}
where $\sigma_0^2 = \langle V_{\mathcal{S}\mathcal{E}}^2 \rangle_{\mathcal{S}_0+\mathcal{E}_0} - \langle V_{\mathcal{S}\mathcal{E}} \rangle_{\mathcal{S}_0+\mathcal{E}_0}^2$ is the variance of the interaction energy in the reference setup, where the system and environment are uncoupled. 
Using the identity between the target and reference states provided by Eq.~(\ref{FEP-like}), and substituting the results from Eq.~(\ref{sigma}) and Eq.~(\ref{sigma0}), we find  
\begin{equation}
\frac{e^{-\beta\mu}}{e^{-\beta\mu_0}} = \chi^2\left(P_\beta \parallel P_{\mathcal{S}_0}\right) + 1, 
\label{mumu0}
\end{equation}  
where $\mu = \langle V_{\mathcal{S}\mathcal{E}} \rangle_{\mathcal{S}+\mathcal{E}}$ and $\mu_0 = \langle V_{\mathcal{S}\mathcal{E}} \rangle_{\mathcal{S}_0+\mathcal{E}_0}$ are the mean interaction when the system and environment are fully coupled and uncoupled, respectively.
Eqs.~(\ref{sigma}) and (\ref{sigma0}) explicitly connect the chi-squared divergence to the variance of $V_{\mathcal{SE}}$. 
These relations imply that the variance of $V_{\mathcal{SE}}$ is invariant upon coupling (i.e. $\sigma^2 = \sigma_0^2$) which allowed us to derive Eq.~(\ref{mumu0}). This relation quantifies how the mean interaction $\mu$ shifts upon coupling in terms of $\chi^2$.
For a better understanding of this main result, it is rewarding to compare it with the Jarzynski equality (JE), which is one of the key relations in non-equilibrium thermodynamics. 
Eq.~(\ref{General}) describes systems that are strongly coupled to their environment in a different way than the JE. 
In terms of importance, it gives a complementary view of how free energy changes. 
The JE,  
$\left\langle e^{-\beta W} \right\rangle = e^{-\beta \Delta F}$ 
was originally derived under the assumption that the system and its environment are only weakly coupled \cite{jarzynski1997nonequilibrium}.
This led to questions about whether it still holds when the coupling is strong \cite{cohen2004note}. 
Later, experimental \cite{ collin2005verification,an2015experimental,liu2023experimental} and theoretical  studies showed that the equality remains valid in the strong coupling regime, as long as the free energy is defined using the HMF framework \cite{jarzynski2004nonequilibrium,campisi2009fluctuation,seifert2016first,jarzynski2017stochastic,jarzynski2012equalities}.  
Eq.~(\ref{General}) includes the effect of coupling from the beginning and, alongside JE, provides another way to describe free energy differences. 
With this in mind, we combine or result with the JE and arrive at
\begin{equation}
\frac{\left\langle e^{-\beta \Delta \mathcal{H}_\beta(x)}\right\rangle_\mathcal{S}}{1+\chi^2\left(P_\beta \parallel P_{\mathcal{S}_0}\right)} = \langle e^{-\beta W}\rangle.
\label{work}
\end{equation} 
To simplify the discussion and get a clearer picture, we return to the earlier assumptions: the condition $\text{Var}_{\mathcal{E}_0} (e^{-\beta V_{\mathcal{SE}}}) = 0$ holds, and the interaction energy $V_{\mathcal{SE}} $ follows a Gaussian distribution.  
Under these assumptions, we rewrite Eq.~(\ref{work}) to yield
\begin{align}
\frac{\Bigr\langle\exp\left( -\beta {V}_{\mathcal{S}\mathcal{E}} \right)\Bigr\rangle_{\mathcal{S+E}}}{1+\chi^2\left(P_\beta \parallel P_{\mathcal{S}_0}\right)} = \langle e^{-\beta W}\rangle. \label{fre}
\end{align}
Now, finally with Eq.~(\ref{sigma}) and Eq.~(\ref{fre}), we obtain
\begin{align}
e^{-\beta \mu}=\frac{\Bigr\langle e^{-\beta W}\Bigr\rangle}{(\chi^2\left(P_\beta \parallel P_{\mathcal{S}_0}\right)+1)^{-0.5} }\label{mu},
\end{align}
or alternatively by combining Eq.~(\ref{mumu0}) and Eq.~(\ref{mu}), we can write
\begin{align}
e^{-\beta \mu_0}=\frac{\Bigr\langle e^{-\beta W}\Bigr\rangle}{(\chi^2\left(P_\beta \parallel P_{\mathcal{S}_0}\right)+1)^{0.5} }.
\label{mu0}
\end{align}
What makes Eqs.~(\ref{mu}) and (\ref{mu0}) particularly interesting is that they connect a quantity that is not directly accessible --- the average interaction energy --- with two observable quantities: the average non-equilibrium work and the shift in probability distributions. 
In our companion study (see \cite{PRE}), we derive exact expressions for both the mean and the fluctuation of the interaction energy. 
These are written in terms of the single-particle density $\rho(\bullet)$ and multi-body structural functions, such as the pair correlation function $g^{(2)}(\bullet,\bullet)$ and higher-order analogues. 
As is evident from Eqs.~(\ref{sigma}), (\ref{sigma0}), (\ref{mu}), (\ref{mu0}), and (\ref{mumu0}), combining the framework presented here with the exact expressions for mean and variance of the interaction energy distribution function established in the accompanying article \cite{PRE} reveals a layered picture: both local structure and nonlocal correlations contribute meaningfully to the behavior of the strongly coupled system. 
The resulting formulation is self-consistent and physically transparent, offering a direct route toward understanding the thermodynamics of strongly coupled open systems.\\
\textit{Conclusion}: In this work, we developed a general framework for thermodynamic potentials that depends explicitly on the full probability distribution of the interaction energy $V_{\mathcal{S}\mathcal{E}}$, rather than relying on the HMF formalism \cite{talkner2016open,talkner2020colloquium,campisi2009fluctuation}. We showed that when the variance $\operatorname{Var}_{\mathcal{E}_0}(e^{-\beta V_{\mathcal{S}\mathcal{E}}})$ vanishes, thermodynamic observables are uniquely determined by the distributions $P_\beta(x_{\mathcal{S}})$ and $P(V_{\mathcal{S}\mathcal{E}})$, thereby resolving the ambiguity of $\mathcal{H}^*_\beta(x_{\mathcal{S}})$ and $F^*_{\mathcal{S}}$ in the strong coupling regime \cite{xing2024thermodynamics,talkner2020colloquium}. 
We derived a universal equality that holds at arbitrary coupling strength. A strict inequality then follows, consistent with Jensen’s inequality applied to the Gibbs--Bogoliubov--Feynman bound \cite{reible2023finite}. 
By combining this result with Jarzynski’s equality \cite{jarzynski1997nonequilibrium}, we further established a thermodynamic relation that clarifies how structural features \cite{PRE} of $P(V_{\mathcal{S}\mathcal{E}})$ influence the thermodynamic response of the strongly coupled open system.
\begin{acknowledgments} We gratefully acknowledge ﬁnancial support by the DFG
under Germany’s Excellence Strategy EXC 2089/1-390776260
(e-conversion).
\end{acknowledgments}

\bibliography{apssamp}

\end{document}